\documentclass[11pt,oneside]{article}
\usepackage{a4wide}
\usepackage{mathrsfs}
\usepackage{latexsym,bm}
\usepackage{graphicx}
\usepackage{indentfirst}
\usepackage{slashed}
\usepackage{amsmath}
\usepackage{amssymb}
\usepackage{color}
\usepackage{hyperref}
\usepackage{epsfig}
\usepackage[titletoc]{appendix}
\usepackage{multirow}%
\usepackage{rotating}
\usepackage{cite}
\usepackage{ulem} 
\usepackage{extpfeil}
\usepackage[top=2.9cm,bottom=2.5cm,left=2.8cm,right=3cm]{geometry}
\usepackage{tabularx}

\newcommand{\jpsi}{J/\psi}

\newcommand{\bra}{\langle}
\newcommand{\ket}{\rangle}

\newcommand{\rxt}{{\rm R}\chi{\rm T}}

\begin{document}
\thispagestyle{empty}
\title{
\Large \bf Effective-Lagrangian study of $\psi'(J/\psi) \to VP$ and the insights into  $\rho\pi$ puzzle } 
\author{\small Lin-Wan Yan$^{a}$,\, Yun-Hua Chen$^{b}$,\, Chun-Gui Duan$^{a}$,\, Zhi-Hui Guo$^{a}$ \\[0.3em]
{ \small\it ${}^a$  Department of Physics and Hebei Key Laboratory of Photophysics Research and Application, } \\
{\small\it Hebei Normal University,  Shijiazhuang 050024, China}\\[0.1em]
{\small\it${}^b$  School of Mathematics and Physics,}\\
{\small\it University of Science and Technology Beijing, Beijing 100083, China } \\[0.1em]
}
\date{}

%

\maketitle
\begin{abstract}
Within the effective Lagrangian approach, we carry out a unified study of the $J/\psi (\psi')\to VP$, $\jpsi \to P\gamma$ and relevant radiative decays of light-flavor hadrons. Large amount of experimental data, including the various decay widths and electromagnetic form factors, are fitted to constrain the numerous hadron couplings. Relative strengths between the strong and electromagnetic interactions are revealed in the $J/\psi \to VP$ and $\psi' \to VP$ processes. The effect from the strong interaction is found to dominate in the $J/\psi \to \rho\pi$ decay, while the electromagnetic interaction turns out to be the dominant effect in $\psi' \to \rho\pi$ decay, which provides an explanation to the $\rho\pi$ puzzle. For the $J/\psi\to K^{*}\bar{K}+\bar{K}^{*}K$ and $\psi' \to K^{*}\bar{K}+\bar{K}^{*}K$, the former process is dominated by the strong interactions, and the effects from the electromagnetic parts are found to be comparable with those of strong interactions in the latter process. Different $SU(3)$ breaking effects from the electromagnetic parts appear in the charged and neutral channels for the $\psi' \to K^{*}\bar{K}+\bar{K}^{*}K$ processes explain the rather different ratios between $B(\psi'\to K^{*+}K^{-}+K^{*-}K^{+})/B(\jpsi\to K^{*+}K^{-}+K^{*-}K^{+})$ and $B(\psi'\to K^{*0}\bar{K}^{0}+\bar{K}^{*0}K^{0})/B(\jpsi\to K^{*0}\bar{K}^{0}+\bar{K}^{*0}K^{0})$. 
\end{abstract}

\section{Introduction}

The strong suppression of the branching ratios of the $\psi'\to \rho \pi$ and $\psi'\to K^*\bar{K}+c.c.$ processes, relative to those of the corresponding decay channels of the $\jpsi$, has been a long-standing puzzle in charmonium physics. The annihilation of the $c\bar{c}$ into three gluons is usually assumed to be the dominant mechanism that rules the decays of the $\jpsi$ and $\psi'$ to the light-flavor hadrons~\cite{Appelquist:1974yr,Appelquist:1974zd,DeRujula:1974rkb}. The annihilation amplitudes of the latter processes and also the decays to the lepton pairs are proportional to the wave functions of the $S$-wave charmonium states $\jpsi$ and $\psi'$ at the origin. As a result, the branching ratios with the light-flavor-hadron ($h$) decays of the $\psi'$ and $\jpsi$ can be predicted by their leptonic decay widths~\cite{ParticleDataGroup:2022pth}, $i.e.$, 
\begin{eqnarray} \label{eq.qh0}
Q_{h}\equiv\frac{B(\psi'\to h)}{B(\jpsi\to h)} =\frac{B(\psi'\to e^+e^-)}{B(\jpsi\to e^+e^-)} = (13.3\pm0.4)\% \,. 
\end{eqnarray}
However, according to the recent PDG averages~\cite{ParticleDataGroup:2022pth}, the ratio of $Q_{\rho\pi}$,  
\begin{eqnarray} 
\frac{B(\psi'\to \rho\pi)}{B(\jpsi\to \rho\pi)} =(0.2\pm0.1)\% \,, 
\end{eqnarray}
and the various ratios of $Q_{K^*\bar{K}}$,
\begin{eqnarray}\label{eq.qkk} 
\frac{B(\psi'\to K^{*+}K^{-}+c.c.)}{B(\jpsi\to K^{*+}K^{-}+c.c.)} &=&(0.5_{-0.1}^{+0.2})\% \,, \nonumber \\
\frac{B(\psi'\to K^{*0}\bar{K}^{0}+c.c.)}{B(\jpsi\to K^{*0}\bar{K}^{0}+c.c.)} &=&(2.6_{-0.7}^{+0.8})\% \,,\nonumber \\
\frac{B(\psi'\to K^*\bar{K}+c.c.)}{B(\jpsi\to K^*\bar{K}+c.c.)} &=&(1.4_{-0.4}^{+0.5})\% \,, 
\end{eqnarray}
are drastically different from the prediction in Eq.~\eqref{eq.qh0}. These contradictions are generally referred as the $\rho\pi$ puzzle, which was first established in Ref.~\cite{Franklin:1983ve} four decades ago. Tremendous efforts have been made to address this problem, including the proposal of a vector glueball near the $\jpsi$ mass~\cite{Suzuki:2000yq}, higher Fock components in the charmonium states~\cite{Chen:1998ma}, the intrinsic charm portions in the light-flavor vector $\rho$~\cite{Brodsky:1997fj}, the nodes in the wave functions~\cite{Pinsky:1989ue}, the meson mixing mechanisms~\cite{Feldmann:2000hs}, the final-state interactions~\cite{Li:1996yn,Zhao:2010mm,Wang:2012mf,Gerard:2013gya}. Another important issue in those decays is the sizable $SU(3)$ breaking effects in the charged and neutral $K^*\bar{K}$ decays of the $\jpsi$ and $\psi'$. I.e., the strict $SU(3)$ flavor symmetry would give the prediction $Q_{K^{*+}K^-+c.c.}=Q_{K^{*0}\bar{K}^{0}+c.c.}$, which is however severely violated according to the experimental measurements in Eq.~\eqref{eq.qkk}.

In this work, we aim at a unified description of the processes $\jpsi(\psi')\to V P$ and $\gamma^{(*)} P$, with $V$ and $P$ the light-flavor vector and pseudoscalar mesons in order. In such kinds of decay processes, one needs to simultaneously take into account of the single-Okubo-Zweig-Iizuka (OZI) or even the doubly suppressed OZI strong interaction effects, the electromagnetic contributions and the $SU(3)$ flavor symmetry breaking terms. 
The effective Lagrangian approach can provide an excellent framework to properly include all the aforementioned effects. Regarding the well-celebrated OZI rule, a quantitative way to understand such suppression mechanism is the large $N_C$ QCD~\cite{Nc}. In the effective field theory (EFT) approach, the $N_C$ counting order can be directly related to the number of traces in the flavor space~\cite{Gasser:1983yg,Gasser:1984gg}. Typically one additional flavor trace will introduce one more $1/N_C$ suppression order to the EFT operator. The single OZI/double OZI effects can be systematically incorporated via the EFT operators with the proper numbers of flavor traces. Furthermore, the chiral EFT is constructed according to the spontaneous and explicit chiral symmetry breaking patterns of QCD. The $SU(3)$ flavor symmetry breaking effects can be then introduced through the basic building tensors of the EFT involving the small but nonvanishing light-flavor quark masses. Although the chiral power counting scheme based on the momentum expansion is not valid in the massive $\jpsi$ or $\psi'$ decays, the basic building blocks and methodology of the EFT Lagrangians are useful to conveniently take into account all the relevant ingredients describing the $\jpsi(\psi')\to V P$ and $\jpsi\to \gamma^{(*)} P$ processes, including the OZI strong interaction parts, the electromagnetic contributions and the $SU(3)$ flavor symmetry breaking effects. This formalism has been successfully applied to the light-flavor decay processes of $V\to P\gamma^{(*)}$, $e^+e^-\to K^*\bar{K}+c.c.$ and $\jpsi\to VP, P\gamma^{(*)}$ in a series of works in Refs.~\cite{Chen:2012vw,Chen:2013nna,Chen:2014yta}. In this work we push forward the study along the line of this research to address the mysterious $\rho\pi$ puzzle by including similar decay processes of $\psi'$. In addition, we also perform the global analyses of the large amount of updated branching ratios of various decay processes from the PDG~\cite{ParticleDataGroup:2022pth} and the newly measured different decay widths from the BESIII collaboration~\cite{BESIII:2018qzg}. 

This paper is organized as follows. In Sec.~\ref{sec.theo}, we introduce the relevant effective Lagrangians and elaborate the calculations of the decay amplitudes. The global fit to the various experimental data and the phenomenological consequences are analyzed in detail in Sec.~\ref{sec.pheno}. We give the short summary and conclusions in Sec.~\ref{sec.summary}.

\section{Effective Lagrangian and calculations of transition amplitudes}\label{sec.theo}

The primary aim of this work is to study the various decay processes of the $\jpsi$ and $\psi'$ into a light-flavor vector and a light pseudoscalar meson, and the light-flavor meson radiative decays and relevant form factors. Therefore we need to include not only the transition operators between the charmonia and the light-flavor mesons, but also the EFT operators describing the interactions among the light-meson themselves. 
To tightly constrain the free couplings, we simultaneously take into account the experimental data from both the decay processes with only light-flavor mesons and also the processes involving the $\jpsi$ and $\psi'$.

Resonance chiral theory (R$\chi$T)~\cite{Ecker:1988te} provides a reliable framework to study the interactions of the light-flavor resonances and the light pseudoscalar mesons ($\pi,K,\eta$), the latter of which are treated as the pseudo-Nambu-Goldstone bosons (pNGBs) resulting from the spontaneous symmetry breaking of QCD. As an extension of the chiral perturbation theory ($\chi$PT), $\rxt$ explicitly introduces the heavier degrees of freedom of QCD, i.e., the light-flavor resonances, such as the vectors $\rho,K^*,\omega,\phi$, the axial vectors, scalars, etc, into the chiral Lagrangians, together with the pNGBs and external source fields, like the photons. The $\rxt$ operators are constructed in a chiral covariant way, therefore the physical amplitudes calculated in the $\rxt$ automatically fulfill the requirements of chiral symmetry of QCD in the low energy region. On the other hand, the large $N_C$ expansion of QCD~\cite{ref:largenc} has been widely used as another useful guide to arrange the operators and amplitudes of the $\rxt$~\cite{Cirigliano:2006hb}. In addition, from the large $N_C$ point of view, the QCD $U_A(1)$ anomaly effect, which is considered to be the most responsible factor for the large mass of the physical state $\eta'$, is however $1/N_C$ suppressed. As a result, the $\eta'$ state would become the ninth pNGB both in large $N_C$ and chiral limits. Based on this argument, the nonet of the pNGBs ($\pi,K,\eta,\eta'$) can be systematically included in the effective  Lagrangian~\cite{ref:ua1nc}. We closely follow this  guideline to include the singlet $\eta_0$ state in the R$\chi$T and adopt the general two-mixing-angle formalism to study the physical processes with the $\eta$ and $\eta'$ mesons. 
Next we briefly introduce the relevant $\rxt$ Lagrangians.

In the present work, only the light-flavor vector resonances will be relevant to our study and the minimal interaction operators with the vectors in even-intrinsic-parity sector of the R$\chi$T is given by~\cite{Ecker:1988te}
\begin{eqnarray}\label{eq.lagv2}
\mathscr{L}_{V}^{(2)} &=&\frac{F_V}{2\sqrt{2}} \bra V_{\mu\nu} f_{+}^{\mu\nu} \ket + i\frac{G_V}{\sqrt{2}} \bra V_{\mu\nu} u^{\mu} u^{\nu} \ket\,, 
\end{eqnarray}
where the nonent of the vector resonances is incorporated via the $3\times 3$ matrix 
\begin{equation}\label{defu3v}
V_{\mu\nu}=
 \left( {\begin{array}{ccc}
   {\frac{1}{\sqrt{2}}\rho ^0 +\frac{1}{\sqrt{6}}\omega _8+\frac{1}{\sqrt{3}}\omega _0 } & {\rho^+ } & {K^{\ast+} }  \\
   {\rho^- } & {-\frac{1}{\sqrt{2}}\rho ^0 +\frac{1}{\sqrt{6}}\omega _8+\frac{1}{\sqrt{3}}\omega _0} & {K^{\ast0} }  \\
   { K^{\ast-}} & {\overline{K}^{\ast0} } & {-\frac{2}{\sqrt{6}}\omega_8+\frac{1}{\sqrt{3}}\omega _0 }  \\
\end{array}} \right)_{\mu\nu}\,,
\end{equation}
the basic chiral tensors with the pNGBs and the external source fields are defined as
\begin{eqnarray}\label{defbb}
&&U=u^2=e^{i\frac{\sqrt{2}\Phi}{F}}\,, \quad u_\mu  =
i \big[ u^\dagger (\partial_\mu - i r_\mu) u - u(\partial_\mu - u\ell_\mu) u^\dagger \big]
\,, \quad f_\pm^{\mu\nu} = u F_L^{\mu\nu} u^\dagger  \pm 
u^\dagger F_R^{\mu\nu} u\,,  \nonumber \\ &&
  F^{\mu\nu}_{L(R)}=\partial^{\mu}l(r)^\nu-\partial^{\nu}l(r)^\mu\,, \quad \chi_\pm  = u^\dagger   \chi u^\dagger \pm  u  \chi^\dagger  u \,, \quad \quad \chi=2 B_0 (s + i p)\,
\end{eqnarray}
and the flavor contents of the nonet pNGB matrix read  
\begin{equation}
\Phi=
 \left( {\begin{array}{*{3}c}
   {\frac{1}{\sqrt{2}}\pi ^0 +\frac{1}{\sqrt{6}}\eta _8+\frac{1}{\sqrt{3}}\eta_0 } & {\pi^+ } & {K^+ }  \\
   {\pi^- } & {-\frac{1}{\sqrt{2}}\pi ^0 +\frac{1}{\sqrt{6}}\eta _8+\frac{1}{\sqrt{3}}\eta_0} & {K^0 }  \\
   { K^-} & {\overline{K}^0 } & {-\frac{2}{\sqrt{6}}\eta_8+\frac{1}{\sqrt{3}}\eta_0 }  \\
\end{array}} \right) \,.
\end{equation}
The quark-mass terms are introduced by taking the scalar external source filed $s$ in Eq.~\eqref{defbb} as $s={\rm diag}\{m_u,m_d,m_s\}$. In this work, we will take $m_u=m_d=\hat{m}$ throughout, i.e. neglecting the isospin breaking effects from the strong interaction parts.  The physical vectors $\omega$ and $\phi$ can be well described by assuming the ideal mixing of the octet $\omega_8$ and the singlet $\omega_0$~\cite{ParticleDataGroup:2022pth}
\begin{eqnarray}
\omega_0&=&\sqrt{\frac{2}{3}}\omega-\sqrt{\frac{1}{3}}\phi ,  \nonumber \\
   \omega_8&=&\sqrt{\frac{2}{3}}\phi+\sqrt{\frac{1}{3}}\omega.
\end{eqnarray}
In contrast, the mixing pattern of the $\eta_8$ and $\eta_0$ is more involved. The modern chiral prescription introduces the sophisticated two-mixing-angle scheme~\cite{Leutwyler:1997yr,Kaiser:1998ds} to address the $\eta$-$\eta'$ mixing system
\begin{eqnarray} \label{eq.twoanglesmixing}
\left(
\begin{array}{c}
\eta   \\
\eta' \\
\end{array}
\right) = \frac{1}{F}\left(
                                        \begin{array}{cc}
                                          F_8\, \cos{\theta_8}  & -F_0 \,\sin{\theta_0}  \\
                                           F_8\,\sin{\theta_8} & F_0 \,\cos{\theta_0} \\
                                        \end{array}
    \right)
      \left(
       \begin{array}{c}
       \eta_8   \\
       \eta_0  \\
       \end{array}
        \right)\,,
\end{eqnarray}
where $F_0$ and $F_8$ are the weak decay constants of the singlet and octet axial-vector currents, respectively. The conventional mixing formula with a single mixing angle can be naturally recovered by taking $F_8=F_0=F$ and $\theta_0=\theta_8$ in Eq.\eqref{eq.twoanglesmixing}. Equivalently one can also use the quark-flavor basis to describe the two-mixing-angle formalism  
\begin{eqnarray} \label{eq.twoanglesmixingqs}
\left(
\begin{array}{c}
\eta   \\
\eta' \\
\end{array}
\right) = \frac{1}{F}\left(
                                        \begin{array}{cc}
                                          F_q\, \cos{\theta_q}  & -F_s \,\sin{\theta_s}  \\
                                           F_q\,\sin{\theta_q} & F_s \,\cos{\theta_s} \\
                                        \end{array}
    \right)
      \left(
       \begin{array}{c}
       \eta_q   \\
       \eta_s  \\
       \end{array}
        \right)\,,
\end{eqnarray}
where the quark-flavor contents of the states $\eta_q=(\eta_8+\sqrt{2}\eta_0)/\sqrt{3}$ and $\eta_s=(-\sqrt{2}\eta_8+\eta_0)/\sqrt{3}$ are $(\bar{u}u+\bar{d}d)/\sqrt{2}$ and $\bar{s}s$, respectively. 

The vector resonances in Eq.~\eqref{eq.lagv2} are expressed in terms of the anti-symmetric tensors, instead of the commonly used Proca fields. It is demonstrated in Refs.~\cite{Ecker:1988te,Ecker:1989yg} that it is convenient to use the anti-symmetric tensors to describe the vector resonances in R$\chi$T, since the high energy behaviors of the resulting amplitudes and form factors automatically match the QCD constraints without requiring the inclusion of extra local chiral counter terms in the anti-symmetric tensor formalism. The $\rxt$ Lagrangians in the odd-intrinsic-parity sector comprise two different classes, namely the $VVP$ and $VJP$ types, with $J$ the external sources. Those $\rxt$ operators written in terms of the anti-symmetric tensor fields that are relevant to the $O(p^4)$ chiral low energy constants, are worked out in Ref.~\cite{RuizFemenia:2003hm}, and the relevant $\rxt$ Lagrangians and discussions on the $VVP$ Green functions by explicitly including the dynamical singlet $\eta_0$ state are given in Ref.~\cite{Chen:2012vw}. A more complete basis of the odd-intrinsic-parity $\rxt$ operators that contribute to the $O(p^6)$ chiral low energy constants, is given in Ref.~\cite{Kampf:2011ty}. A proliferation of the unknown resonance couplings arise in the more complete $\rxt$ Lagrangians, as expected. This can hinder one from giving the definite conclusions on the phenomenological discussions~\cite{Miranda:2020wdg}. From the practical point of view, we will work with the $\rxt$ operator basis from Refs.~\cite{RuizFemenia:2003hm,Chen:2012vw} and we believe that the higher order effects from the extra operators in Ref.~\cite{Kampf:2011ty} can be accounted for by the uncertainties of the resonance couplings in the former two references. For the sake of completeness, we give the explicit expressions of the relevant $\rxt$ Lagrangians~\cite{RuizFemenia:2003hm,Chen:2012vw}
\begin{eqnarray}\label{eq.lagvvp}
\mathscr{L}_{VVP}= && d_1 \varepsilon_{\mu\nu\rho\sigma} \langle \{V^{\mu\nu}, V^{\rho\alpha}\} \nabla_{\alpha}u^{\sigma}  \rangle
+ i d_2 \varepsilon_{\mu\nu\rho\sigma} \langle \{V^{\mu\nu}, V^{\rho\sigma}\} \chi_- \rangle + d_3 \varepsilon_{\mu\nu\rho\sigma} \langle \{ \nabla_\alpha V^{\mu\nu}, V^{\rho\alpha}\} u^{\sigma}  \rangle
\nonumber \\ && 
+ d_4 \varepsilon_{\mu\nu\rho\sigma} \langle \{ \nabla^\sigma V^{\mu\nu}, V^{\rho\alpha}\} u_{\alpha}  \rangle -i d_5M_V^2\sqrt{\frac{2}{3}}\varepsilon_{\mu\nu\rho\sigma}\langle
 V^{\mu\nu}V^{\rho\sigma}\rangle \ln(\det u) \,,
\end{eqnarray}
and
\begin{eqnarray}\label{eq.lagvjp}
\mathscr{L}_{VJP}=&& \frac{c_1}{M_V} \varepsilon_{\mu\nu\rho\sigma} \langle \{ V^{\mu\nu}, f_+^{\rho\alpha} \} \nabla_{\alpha}u^{\sigma}  \rangle
+\frac{c_2}{M_V} \varepsilon_{\mu\nu\rho\sigma} \langle \{ V^{\mu\alpha}, f_+^{\rho\sigma} \} \nabla_{\alpha}u^{\nu}  \rangle + \frac{i c_3}{M_V} \varepsilon_{\mu\nu\rho\sigma} \langle \{ V^{\mu\nu}, f_+^{\rho\sigma}\} \chi_- \rangle
 \nonumber \\ && 
+\frac{i c_4}{M_V} \varepsilon_{\mu\nu\rho\sigma} \langle V^{\mu\nu} [ f_-^{\rho\sigma}, \chi_+ ] \rangle
\nonumber +  \frac{c_5}{M_V} \varepsilon_{\mu\nu\rho\sigma} \langle \{ \nabla_{\alpha} V^{\mu\nu}, f_+^{\rho\alpha}\} u^{\sigma}  \rangle
+ \frac{c_6}{M_V} \varepsilon_{\mu\nu\rho\sigma} \langle \{ \nabla_{\alpha} V^{\mu\alpha}, f_+^{\rho\sigma}\} u^{\nu}  \rangle
\nonumber \\ &&+ \frac{c_7}{M_V} \varepsilon_{\mu\nu\rho\sigma} \langle \{ \nabla^{\sigma} V^{\mu\nu}, f_+^{\rho\alpha}\} u_{\alpha}  \rangle - ic_8M_V \sqrt{\frac{2}{3}}\varepsilon_{\mu\nu\rho\sigma}\langle
V^{\mu\nu}\tilde{f}_+^{\rho\sigma}\rangle \ln(\det u) \,,
\end{eqnarray}
where the covariant derivative acting on the chiral field $X$ is given by 
\begin{eqnarray}
\nabla_{\mu}X=\partial_{\mu}X+[\Gamma_{\mu},X]\,, \qquad 
\Gamma_{\mu}=\frac{1}{2}\big[ u^{+}(\partial
_\mu-ir_\mu)u+u(\partial
_\mu-il_\mu)u^{+} \big]\,.
\end{eqnarray}

As previously mentioned in the Introduction, both the strong and electromagnetic interactions can be important in the $\jpsi(\psi')\to VP$ processes. The effects from the strong interactions are taken into account by the direct $\jpsi(\psi') VP$ transition operators~\cite{Chen:2012vw}
\begin{eqnarray}\label{eq.lagpsivp}
\mathscr{L}_{\psi(\psi') VP}=&& M_{\psi(\psi')} h^{(\prime)}_1 \varepsilon_{\mu\nu\rho\sigma} \psi^{(\prime)\mu}\bra u^\nu V^{\rho\sigma}\ket
+\frac{1}{M_{\psi(\psi')}} h^{(\prime)}_2 \varepsilon_{\mu\nu\rho\sigma}
\psi^{(\prime)\mu}\bra \{u^\nu, V^{\rho\sigma}\}\chi_+\ket  \nonumber \\
&& +M_{\psi(\psi')}
h^{(\prime)}_3\varepsilon_{\mu\nu\rho\sigma}\psi^{(\prime)\mu}\bra u^\nu\ket \bra V^{\rho\sigma}\ket \,,
\end{eqnarray}
where the couplings $h^{(\prime)}_{i=1,2,3}$ corresponding to the $\jpsi$ and $\psi'$ will be separately fitted to the experimental data of the two charmonium states. Two types of EFT operators are introduced to account for the electromagnetic effects, which include the direct $\psi P\gamma$ transition operators
\begin{eqnarray}\label{eq.lagpsigp}
\mathscr{L}_{\psi P\gamma }=g_1 \varepsilon_{\mu\nu\rho\sigma} \psi^\mu\bra u^\nu f_+^{\rho\sigma}\ket
+\frac{1}{M^2_\psi} g_2 \varepsilon_{\mu\nu\rho\sigma}
\psi^\mu\bra \{u^\nu,
f_+^{\rho\sigma}\}\chi_+\ket \,,
\end{eqnarray}
and the conversion vertex of the charmonium and the photon 
\begin{eqnarray}\label{eq.lagpsipho}
\mathscr{L}_{\psi\gamma}=\frac{-1}{2\sqrt{2}}\frac{f_\psi}{M_\psi}\bra \hat{\psi}_{\mu\nu} f_+^{\mu\nu}\ket \,,
\end{eqnarray}
being $\hat{\psi}_{\mu\nu}=\partial_\mu \psi^\nu-\partial_\nu\psi^\mu$. The values of the couplings $g_1$, $g_2$ and $f_\psi$ are different for the $\jpsi$ and $\psi'$ and they will be determined by the relevant experimental data. Different powers of the $M_{\psi(\psi')}$ are introduced in Eqs.~\eqref{eq.lagpsivp}-\eqref{eq.lagpsipho}, so that the couplings appearing in those Lagrangians are dimensionless.

It is found~\cite{Chao:1990im,Chen:2014yta} that the $\jpsi \to \eta^{(')} \gamma^{(*)}$ amplitudes are dominated by the $\eta_c$ mediating diagrams, i.e., via the $\jpsi \to \eta_c \gamma^{(*)}\to  \eta^{(')} \gamma^{(*)}$ intermediate processes. The decay amplitude of the $\psi\to \eta^{(')}\gamma^{(*)}$ can be written as
\begin{eqnarray}\label{eq.etacmixing}
\mathcal{M}_{\psi\to \eta^{(')}\gamma^{\ast}}^{mixing}= 
e\, \varepsilon_{\mu\nu\rho\sigma} \epsilon_\psi^\mu
\epsilon_{\gamma^{\ast}}^\nu q^\rho k^\sigma \,
 \lambda_{\eta_c\eta^{(')}} \, g_{\psi \eta_c\gamma^{*}}(s) \, e^{i\delta_P}\,, 
\end{eqnarray}
being  $P=\eta,\eta'$, where the electromagnetic transition form factor between the $\psi$ and $\eta_c$ takes form~\cite{Dudek:2006ej,Dudek:2009kk,Chen:2011kpa}
\begin{eqnarray}\label{eq.eqpsietacg}  
g_{\psi \eta_c\gamma^{*}}(s) =g_{\psi \eta_c\gamma^{*}}(0)e^{\frac{s}{16\beta^2}} \,. 
\end{eqnarray}
For the mixing parameters $\lambda_{\eta_c\eta^{(')}}$ between the $\eta_c$ and $\eta^{(')}$ states, we take the determinations $\lambda_{\eta_c\eta}=-4.6\times
10^{-3}$ and $\lambda_{\eta_c\eta^{\prime}}=-1.2\times10^{-2}$ from Ref.~\cite{Chao:1990im}. The phenomenological phase factors $\delta_{\eta^{(')}}$ in front of the $\eta_c$ mediating diagrams need to be separately fitted to the data of the $\jpsi$ . 

\begin{figure}[htbp]
\centering
\includegraphics[angle=0, width=0.8\textwidth]{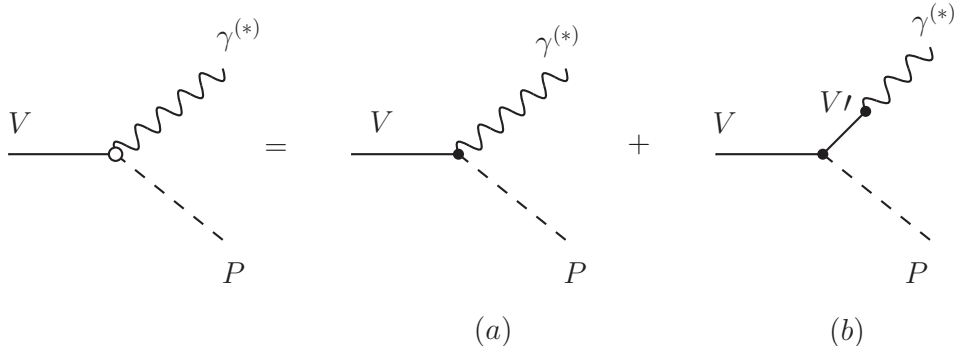}
\caption{ Diagrams relevant to the $V\rightarrow P \gamma^{(*)}$
processes: (a) direct type and (b) indirect type.  }\label{fig.vpg}
\end{figure}

\begin{figure}[htbp]
\centering
\includegraphics[angle=0, width=0.8\textwidth]{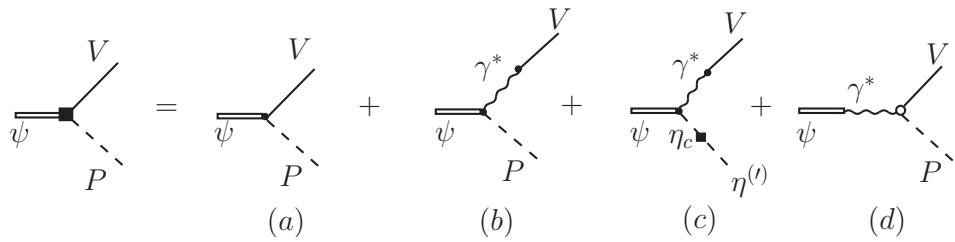}
\caption{ Feynman diagrams for the processes $J/\psi \rightarrow
VP$. The notations of the solid square in diagram (c) and the open circle in diagram (d) are explained in the text. 
}\label{fig.psivp}
\end{figure}

\begin{figure}[h]
\includegraphics[angle=0, width=0.8\textwidth]{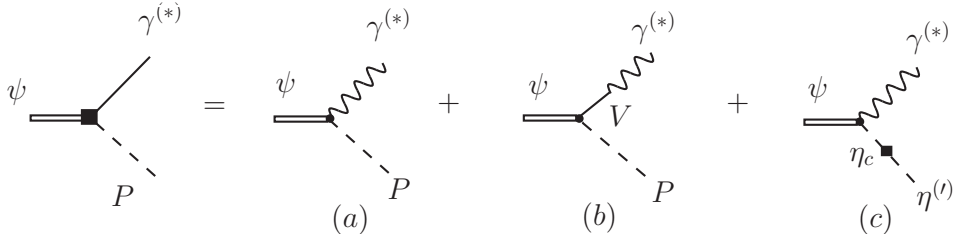}
\caption{ Feynman diagrams for the processes $J/\psi \rightarrow P
\gamma^{(*)}$ }\label{fig.psipg}
\end{figure}

The various Feynman diagrams relevant to our study are illustrated in Figs.\ref{fig.vpg}, \ref{fig.psivp} and \ref{fig.psipg}. To be more specific, the diagrams in Fig.~\ref{fig.vpg} contribute to the light-flavor processes $V\to P\gamma^{(*)}$ and $P\to V \gamma^{(*)}$. The amplitudes of the $\jpsi(\psi')\to VP$ and $\jpsi \to P\gamma^{(*)}$ receive contributions from the diagrams in Figs.~\ref{fig.psivp} and \ref{fig.psipg}, respectively. The formulas relevant to the $V\to P\gamma^{(*)}$ and $P\to V \gamma^{(*)}$ processes are worked out in Ref.~\cite{Chen:2012vw}, and the expressions of the $\jpsi\to VP, P\gamma^{(*)}$ amplitudes are calculated in Ref.~\cite{Chen:2014yta}. The corresponding decay amplitudes of the $\psi'$ state share similar expressions as those involving $\jpsi$, with obvious replacements of the resonance couplings. Nevertheless, for the sake of completeness and to set up the notations, we further elaborate the amplitudes of the processes of $\psi'\to VP$ . 

For the $\psi'\to VP$ decay, the first diagram (a) in Fig.~\ref{fig.psivp} denotes the contributions from the strong interactions, i.e., from the Lagrangians in Eq.~\eqref{eq.lagpsivp}. Other diagrams in Fig.~\ref{fig.psivp} correspond to the electromagnetic effects. The $\psi'\to VP$ amplitude can be written as 
\begin{eqnarray}\label{eq.defgvp}
\mathcal{M}_{\psi'\to VP}=  \varepsilon_{\mu\nu\rho\sigma}
\epsilon_{\psi'}^\mu \epsilon_V^\nu q^\rho k^\sigma  G_{\psi' \to VP}\,,
\end{eqnarray} 
where the polarization vectors of the $\psi'$ and $V$ are given by $\epsilon_{\psi'}^\mu$ and $\epsilon_V^\nu$, $q$ and $k$ stand for the four-momentum of the $\psi'$ and $V$, respectively. The effective couplings $G_{\psi' \to VP}$ include various contributions from the individual diagram of Fig.~\ref{fig.psivp}. The explicit expressions of $G_{\psi'\to VP}$ for the various processes are given in Appendix (\ref{appendix.FFpsip}). The decay widths of $\psi' \to VP$ read 
\begin{eqnarray}\label{widthpsipVP}
\Gamma(\psi' \rightarrow VP)=\frac{1}{96\pi
M_{\psi'}^3} {\lambda(M_{\psi'},M_V,m_P)}^{\frac{3}{2}} \,\big|G_{\psi'\rightarrow VP}\big|^2\,,
\end{eqnarray}
with the K\"all\'en function $\lambda(x,y,z)=x^2+y^2+z^2-2xy-2xz-2yz$.

Similarly, the corresponding amplitude of the radiative process $\jpsi(q) \to \gamma^\ast(k) P(q-k)$ can be given in terms of one effective coupling as well 
\begin{eqnarray}\label{defgpg}
\mathcal{M}_{\psi\to P\gamma^\ast}=  e\,
\varepsilon_{\mu\nu\rho\sigma} \epsilon_\psi^\mu
\epsilon_{\gamma^\ast}^\nu q^\rho k^\sigma
 G_{\psi \to P\gamma^\ast}(s)\,,
\end{eqnarray}
with $s=k^2$. The effective coupling $G_{\psi \to P\gamma^\ast}$ can receive contributions from all the diagrams in Fig.~\ref{fig.psipg}. The explicit expressions are given in Ref.~\cite{Chen:2014yta}. The formula of the decay width of the $\jpsi \to P\gamma$ process finds it form 	
\begin{eqnarray}
\Gamma_{\psi\rightarrow
P\gamma}=\frac{1}{3}\alpha\left(\frac{M_\psi^2-M_P^2}{2M_\psi}\right)^3|G_{\psi\rightarrow
P\gamma^{*}}(0)|^2\,.
\end{eqnarray}
The expression of the width for the Dalitz decay process $\jpsi \to P\gamma^\ast \to P l^+ l^-$ is given by 
\begin{eqnarray}\label{gammapsipll}
\Gamma_{ \psi \to P l^+ l^-}= \int^{(M_\psi-m_P)^2}_{4m_l^2}
\frac{\alpha^2(2 m_l^2+s)}{72 M_\psi^3 \pi
s^3}\sqrt{s(s-4m_l^2)} \, \big[\lambda(s,M_\psi,m_P)]^{\frac{3}{2}} \,
|G_{\psi \to P\gamma^\ast}(s)|^2 d s\,.
\end{eqnarray}

\section{Comprehensive fits and phenomenological discussions}\label{sec.pheno}

Compared to the previous studies in Refs.~\cite{Chen:2012vw,Chen:2014yta}, we incorporate in this work the data from the various $\psi'\to VP$ decays, apart from other types of data from the $V\to P\gamma^{(*)}$, $P\to V \gamma^{(*)}$ and $\jpsi\to VP, P\gamma^{(*)}$ processes, to perform a comprehensive fit, so that the $\rho\pi$ puzzle can be addressed. In addition, we update numerous types of data according to the most recent PDG averages~\cite{ParticleDataGroup:2022pth}, and timely revise the determinations of the resonance couplings. 

In total, we include 135 data points from several different types of processes in the comprehensive fit. To be more specific, the data from the pure light-flavor processes amount to 70, and they consist of both the decay widths, such as those of the $\omega\to\eta\gamma$, $\eta'\to\omega\gamma$, $\eta\to \gamma\gamma$, etc, and the form factors of the $\phi\to\eta\gamma^*$ and $\eta^{(')}\to\gamma\gamma^*$. For the data related to the $\jpsi$, they include all the available widths of the $\jpsi\to VP$, $P\gamma$ and $P e^+e^-$ from the PDG~\cite{ParticleDataGroup:2022pth}, and also the recent BESIII measurements of the invariant-mass distributions of the lepton pairs in the transition of $\jpsi \to \eta \gamma^\ast$~\cite{BESIII:2018qzg}. Regarding the data of the $\psi'$, we will include in the joint fit all the available widths of the $\psi'\to VP$ processes from PDG~\cite{ParticleDataGroup:2022pth}.

An efficient way to reduce the number of unknown couplings in the $\rxt$ is to impose the high energy constraints dictated by QCD to the various form factors and Green functions calculated from the $\rxt$ Lagrangians in Eqs.~\eqref{eq.lagv2}, \eqref{eq.lagvvp} and \eqref{eq.lagvjp}. Furthermore, the high energy behaviors of the resulting amplitudes after imposing such constraints will mimic the properties as predicted by QCD. Following the previous discussions in Refs.~\cite{RuizFemenia:2003hm,Chen:2012vw,Chen:2013nna,Chen:2014yta,Guo:2010dv,Roig:2013baa,Chen:2022nxm}, we take the following high energy constraints on the various couplings 
\begin{eqnarray} \label{eq.he-ope-c8}  
4c_3+c_1=0\,, & \quad
c_1-c_2+c_5=0\,, & \quad 
c_5-c_6=\frac{N_C}{64\pi^2}\frac{M_V}{\sqrt{2}F_V}\,, \nonumber\\
d_1+8d_2 -d_3 = \frac{F^2}{8F_V^2}\,, & \quad  d_3=-\frac{N_C}{64\pi^2} \frac{M_V^2}{F_V^2}\,, & \quad
c_8 =  -\frac{ \sqrt2 M_0^2}{\sqrt3 M_V^2} c_1 \,,
\end{eqnarray}
where the pion weak decay constant takes the normalization $F=92.4$~MeV  throughout, the $U_A(1)$ anomaly parameter is set to be $M_0=900$~MeV~\cite{Guo:2011pa,T. Feldmann:2000IJMPA}, the chiral-limit mass of the lowest vector resonance multiplet is fixed at $M_V=M_\rho=775$~MeV and the vector-photon transition coupling $F_V$ will be fitted. 
By taking into account the the leptonic widths of the $\jpsi$ and $\psi'$, we can determine the charmonium-photon transition coupling $f_{\jpsi(\psi')}$ in Eq.~\eqref{eq.lagpsipho}, whose explicit values are found to be 
\begin{eqnarray}
f_{\jpsi} =293.8 \pm 3.5~ {\rm MeV}\,, \qquad f_{\psi'}= 208.1 \pm 5.1~ {\rm MeV}\,. 
\end{eqnarray}

We are then left with 23 undetermined parameters, including the four $\eta$-$\eta'$ mixing parameters introduced in Eq.~\eqref{eq.twoanglesmixing}, four couplings $F_V$, $c_3, c_4$ and $d_2$ that emerge from the light-flavor resonance interactions, nine parameters exclusively entering in the $\jpsi$ decays and six parameters that are dedicated to the $\psi'$ processes. The couplings that describe the interactions of the light-flavor resonances will also enter in the charmonia decays. Therefore the joint fits by simultaneously including the relevant data of the light-flavor mesons, the data from the $\jpsi\to VP, P\gamma^{(*)}$ and the $\psi'$ ones, will obviously give more stringent constraints on the couplings than the situation by including just one of these data sets. Furthermore, such  comprehensive studies in a unified framework are also expected to give a further insight into $\rho\pi$ puzzle elaborated in the Introduction.

\begin{table}[htbp]
\centering
\begin{scriptsize}
\begin{tabular}{ c c c c }
\hline\hline
$F_8$ &$(1.41\pm 0.02)F_\pi$  & $F_0$ &$(1.36\pm 0.03)F_\pi$  \\ 
$\theta_8$ &$(-24.3\pm 0.4)^\circ$ & $\theta_0$&$(-12.8\pm0.5)^\circ$  \\
$F_V$&$139.04\pm1.72$ & $c_3$&$0.0046\pm0.0003$\\
$c_4$&$-0.0014\pm0.0001$ & $d_2$&$0.100\pm0.008$\\
$h_1$&$(-2.35\pm0.06)\times{10}^{-5}$ & $h_2$&(-3.08$\pm$0.60)$\times{10}^{-5}$ \\
$h_3$&(3.39$\pm$0.22)$\times{10}^{-6}$ & $g_1$&(-2.40$\pm$0.06)$\times{10}^{-5}$\\
$g_2$&(-2.23$\pm$0.48)$\times{10}^{-4}$ & $r_1$&0.40$\pm$0.04\\
$h^\prime_1$&(0.33$\pm$0.23)$\times{10}^{-6}$& 
$h^\prime_2$&(-4.01$\pm$0.32)$\times{10}^{-5}$\\
$h^\prime_3$&(0.85$\pm$0.47)$\times{10}^{-6}$& 
$g^\prime_1$&(-1.70$\pm$0.47)$\times{10}^{-4}$\\
$g^\prime_2$&(0.18$\pm$0.95)$\times{10}^{-3}$&
$\delta_{\eta}$ & $(117.12\pm3.81)^\circ$ \\
$\delta_{\eta'}$ & $(50.03\pm16.01)^\circ$ & 
$\beta$&$ 512.86\pm7.36$~MeV \\
$\beta'$&$112.97\pm0.98$~MeV  &
$F_q^{(*)}$ & (1.24$\pm$0.02)$F_\pi$ \\
$F_s^{(*)}$  & (1.52$\pm$0.02)$F_\pi$ & 
$\theta_q^{(*)}$&$(37.3\pm0.7)^\circ$ \\
$\theta_s^{(*)}$&$(35.1\pm0.4)^\circ$ & 
$\chi^2/d.o.f$& 157.25/(135-23)=1.40 \\
\hline\hline
\end{tabular}
\end{scriptsize}
\caption{ Parameters from the joint fit. The quantities marked with asterisk are predictions, instead of free parameters in the fit.  \label{tab.fitting}   } 
\end{table}

The resulting parameters from the joint fit are given in  Table.~\ref{tab.fitting}. 
The updated parameters related to the light-flavor resonances, the $\jpsi$ decays and the $\eta$-$\eta'$ mixing are well consistent with the previous determinations~\cite{Chen:2012vw,Chen:2013nna,Chen:2014yta} where the data of the $\psi'$ processes are not included in  these studies. For the $\psi'\to P\gamma^{(*)}$ processes, which could receive significant contributions from the $\psi' \to\jpsi P$ transition vertexes, i.e. $\psi' \to \jpsi \eta \to \gamma \eta $~\cite{Q. Zhao:PLB2011}, are not considered in this work. Therefore we will not discuss such kinds of processes here. The relative phases $\delta_{\eta^{(')}}$ of Eq.~\eqref{eq.etacmixing} for the $\eta_c$ mediating effects in the $\psi'\to V \eta^{(')}$ decay processes, are found to be insensitive to our present studies. As a result, the phases of $\delta_{\eta^{(')}}$ in the $\psi'$ decays will be fixed to zero throughout. The previous study in Ref.~\cite{Chen:2012vw} pointed out a strong correlation between the $d_2$ and $d_5$ parameters, and we find that this correlation still holds in our joint fit. The resulting relation turns out to be $d_5=3.57d_2+0.01$. Regarding the four parameters $F_8, F_0, \theta_8$ and $\theta_0$ related with the $\eta$-$\eta'$ mixing, our current determinations of the central values and uncertainties more or less resemble those in Ref.~\cite{Chen:2014yta}. In Ref.~\cite{Chen:2012vw}, only  the data from the light-flavor sector were considered and the resulting $\eta$-$\eta'$ mixing parameters were found to bear large uncertainties. The simultaneous inclusion of the relevant data from the $\jpsi$ and $\psi'$ processes, together with the light-flavor ones, can obviously pin down the uncertainties of the $\eta$-$\eta'$ mixing parameters~\cite{Chen:2014yta,Feldmann:1998vh,Bramon:1997mf,Escribano:2009jti}. In Table~\ref{tab.fitting}, we also give the predictions to the mixing parameters in the quark-flavor basis.

\begin{table}[htbp]
	\centering
	\begin{scriptsize}
		\begin{tabular}{ c c c }
			\hline\hline
 ~~ &Exp&Fit     \\
	\hline
$\Gamma_{\omega \to \pi\gamma}$&$724.78\pm34.64$&$705.65\pm17.40$\\
$\Gamma_{\rho^0 \to \pi^0\gamma}$&$70.08\pm12.37$&$73.23\pm1.81$\\
$\Gamma_{K^{*0} \to K^0\gamma}$&$116.36\pm11.27$&$108.95\pm2.69$\\
$\Gamma_{\omega \to \pi e^-e^+}$&$6.68\pm0.63$&$6.40\pm0.16$\\
$\Gamma_{\omega \to \pi \mu^-\mu^+}$&$1.16\pm0.18$&$0.63\pm0.02$\\
$\Gamma_{\omega \to \eta\gamma}$&$3.91\pm0.41$&$5.30\pm0.11$\\
$\Gamma_{\rho^0 \to \eta\gamma}$&$44.73\pm3.39$&$43.93\pm0.96$\\
$\Gamma_{\phi \to \eta\gamma}$&$55.28\pm1.23$&$55.01\pm1.00$\\
$\Gamma_{\phi \to \eta'\gamma}$&$0.26\pm0.01$&$0.26\pm0.01$\\
$\Gamma_{\eta' \to \omega\gamma}$&$4.74\pm0.29$&$5.05\pm0.18$\\
$\Gamma_{\eta \to \gamma\gamma}$&$0.52\pm0.02$&$0.50\pm0.01$\\
$\Gamma_{\eta' \to \gamma\gamma}$&$4.34\pm0.20$&$3.92\pm0.11$\\
$\Gamma_{\eta \to \gamma e^-e^+}$&${(9.04\pm0.89)}\times10^{-3}$&${(8.32\pm0.23)}\times10^{-3}$\\
$\Gamma_{\eta \to \gamma \mu^-\mu^+}$&${(0.41\pm0.07)}\times10^{-3}$&${(0.39\pm0.01)}\times10^{-3}$\\
$\Gamma_{\eta' \to \gamma \mu^-\mu^+}$&${(2.12\pm0.61)}\times10^{-2}$&${(1.47\pm0.04)}\times10^{-2}$\\
$\Gamma_{\phi \to \eta e^-e^+}$&$0.459\pm0.018$&$0.460\pm0.008$\\			
			\hline\hline
		\end{tabular}
	\end{scriptsize}
	\caption{ The decay widths in units of KeV for the light-flavor hadrons.  \label{tab.light-flavor hadron decays}   } 
\end{table}

\begin{table}[htbp]
	\centering
	\begin{scriptsize}
		\begin{tabular}{ c c c  }
			\hline\hline
			~~ &Exp&Fit     \\
			\hline
$J/\psi \to \rho^0\pi^0$ & $5.6\pm0.7$ & $5.5\pm0.3$\\
$J/\psi \to \rho\pi$ & $16.9\pm1.5$ & $16.2\pm1.0$\\
$J/\psi \to \rho^0\eta$ & $0.193\pm0.023$ & $0.185\pm0.021$\\
$J/\psi \to \rho^0\eta'$ & $0.081\pm0.008$ & $0.080\pm0.007$\\
$J/\psi \to \omega\pi^0$ & $0.45\pm0.05$ & $0.45\pm0.04$\\
$J/\psi \to \omega\eta$ & $1.74\pm0.20$& $1.65\pm0.09$\\
$J/\psi \to \omega\eta'$ & $0.189\pm0.018$ & $0.189\pm0.018$\\
$J/\psi \to \phi\eta$ & $0.74\pm0.08$ & $0.76\pm0.06$\\
$J/\psi \to \phi\eta'$ & $0.46\pm0.05$ & $0.45\pm0.05$\\
$J/\psi \to K^{*+}K^-+c.c.$ & $6.0\pm1.0$& $6.6\pm0.3$\\
$J/\psi \to K^{*0}\bar{K}^0+c.c.$ & $4.2\pm0.4$ & $3.8\pm0.2$\\
$J/\psi \to \pi^0\gamma$ & $0.0356\pm0.0017$ & $0.0341\pm0.0016$\\
$J/\psi \to \eta\gamma$ & $1.085\pm0.018$& $1.085\pm0.013$\\
$J/\psi \to \eta'\gamma$ & $5.25\pm0.07$& $5.35\pm0.04$\\
$J/\psi \to \pi^0e^+e^-$ & ${(0.076\pm0.014)}\times 10^{-2}$& ${(0.129\pm0.004)}\times 10^{-2}$ \\
$J/\psi \to \eta e^+e^-$ & ${(1.42\pm0.08)}\times 10^{-2}$& ${(1.35\pm0.02)}\times 10^{-2}$ \\
$J/\psi \to \eta' e^+e^-$ & ${(6.59\pm0.18)}\times 10^{-2}$& ${(6.08\pm0.05)}\times 10^{-2}$ \\	
			\hline\hline
		\end{tabular}
	\end{scriptsize}
	\caption{ Branching fractions$(\times 10^{-3})$ of the decay processes for $\jpsi$ . \label{tab.Branching fractions for jpsi decays}  } 
\end{table}

\begin{figure}[htbp]
\centering
\includegraphics[angle=0, width=0.8\textwidth]{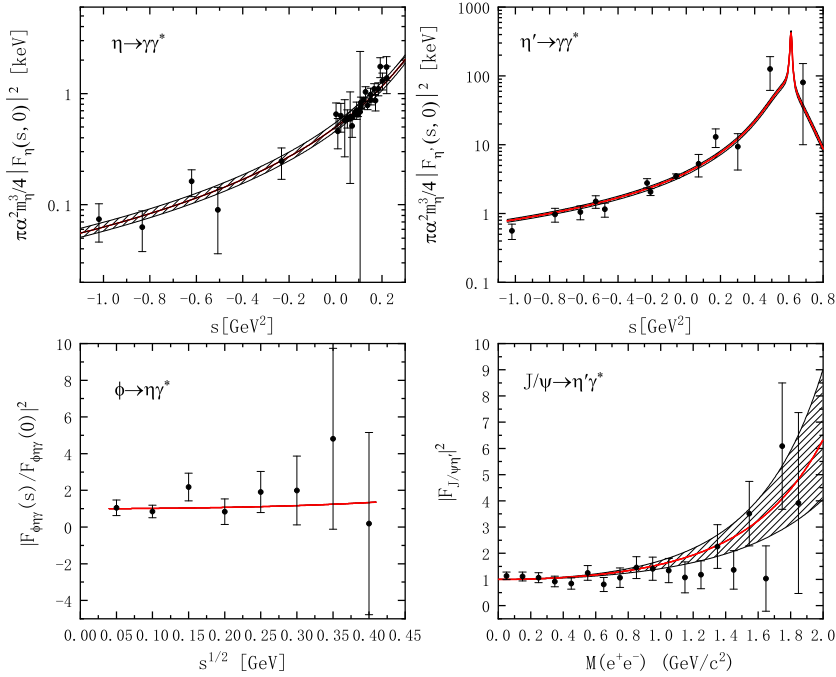}
\caption{ The form factors for the $\eta\to\gamma\gamma^*$, $\eta'\to\gamma\gamma^*$, $\phi\to\eta\gamma^*$ and $\jpsi\to\eta'\gamma^*$. The red solid lines are obtained by taking the central values of the parameters in Table~\ref{tab.fitting}, and the shaded areas correspond to the error bands at 1-$\sigma$ level. The experimental data on the form factors for the $\eta\to\gamma\gamma^*$ , $\eta'\to\gamma\gamma^*$, $\phi\to\eta\gamma^*$ and $\jpsi\to\eta'\gamma^*$ are taken from Refs.~\cite{Dzhelyadin:1980kh,Dzhelyadin:1979za,CELLO:1990klc,Achasov:2000ne,NA60:2009una,TPCTwoGamma:1990dho}, Refs.~\cite{Dzhelyadin:1980kh,Dzhelyadin:1979za,CELLO:1990klc,TPCTwoGamma:1990dho,L3:1997ocz}, Ref.~\cite{Achasov:2000ne} and Ref.~\cite{BESIII:2014dax},respectively. }\label{fig.light hardrons}
\end{figure}

\begin{figure}[htbp]
	\centering
	\includegraphics[angle=0, width=0.95\textwidth]{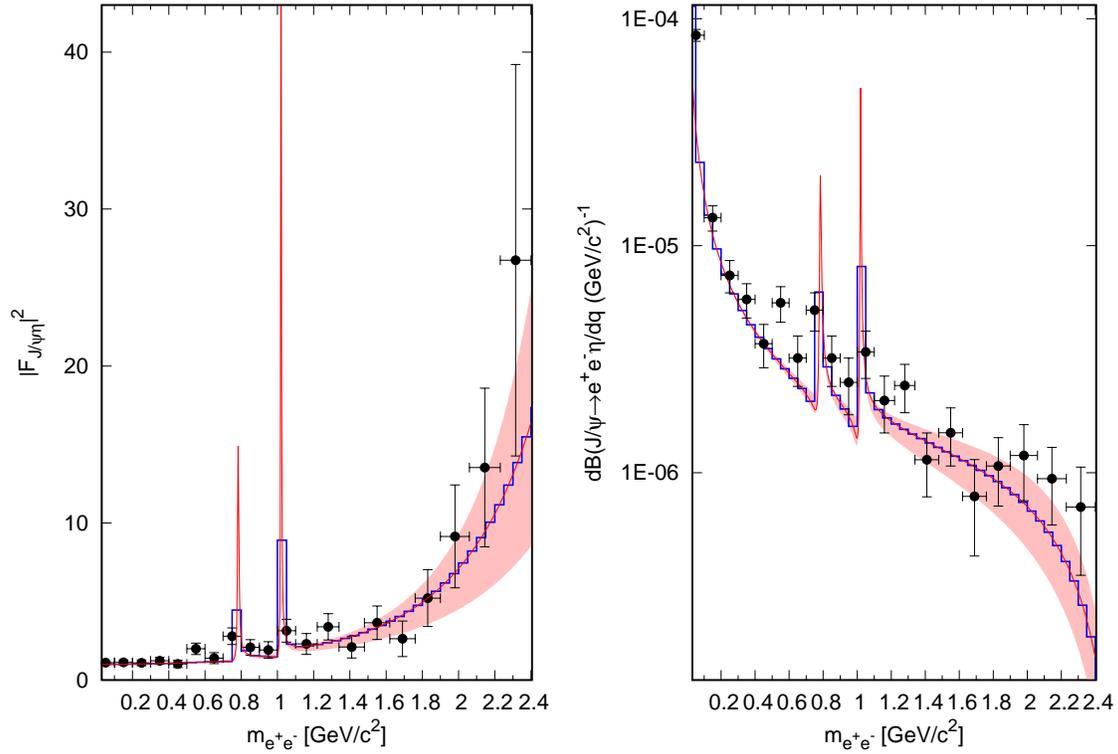}
	\caption{ The form factors and differential branching fractions for the $\jpsi \to \eta e^+e^-$. The experimental data are from the Ref.\cite{BESIII:2018qzg}.  The red solid lines represent the curves with the central values of the parameters in Table.\ref{tab.fitting}, and the shaded areas stand for the error bands. The histograms are obtained by taking different energy bin width at 50~MeV. }\label{fig.dBdqFFpsietaee}
\end{figure}

Generally speaking, the numerous types of data are well reproduced in our comprehensive fit. The comparisons of the various decay widths for the pure light-flavor processes from the revised fit and the updated PDG values are shown in Table.~\ref{tab.light-flavor hadron decays}. Similar comparisons for the partial decay widths of the $\jpsi$ and $\psi'$ are given in Tables.~\ref{tab.Branching fractions for jpsi decays} and \ref{tab.Branching fractions for psip decays}, respectively. The resulting curves of the form factors for the $\eta\to\gamma\gamma^*$, $\eta'\to\gamma\gamma^*$, $\phi\to\eta\gamma^*$ and $\jpsi\to\eta'\gamma^*$ are shown together with the experimental data in Fig.~\ref{fig.light hardrons}. The fitted results of the recent BESIII measurements on the $e^+e^-$ spectra in the $\jpsi\to \eta e^+e^-$ processes are illustrated in Fig.~\ref{fig.dBdqFFpsietaee}. We point out a subtlety about the effects of the light-flavor vector resonances in the $e^+e^-$ spectra. In the $\jpsi\to \eta' e^+e^-$ decays, the light-flavor vectors are removed in the BESIII analysis~\cite{BESIII:2014dax}, and as a result we have also subtracted the contributions from the intermediate light vector exchanges in accord with the experimental setups. This explains the smooth line shapes of the electromagnetic $\jpsi\to \eta'e^+e^-$ transition form factors shown in Fig.~\ref{fig.light hardrons}. Regarding the $\jpsi\to \eta e^+e^-$ process, we keep the effects of the intermediate light-flavor vector resonances, in order to be consistent with the setups of the experimental analyses in Ref.~\cite{BESIII:2018qzg}. It should be stressed that the prominent peaks of the narrow vectors $\omega$ and $\phi$ can be diluted due to the large bin widths of the experimental energy resolutions. To clearly show the influence of the bin widths, we give the histograms by using the energy bin width at 50~MeV. It is evident that the signals of narrow light vector resonances can be obviously enhanced when the energy bin width is reduced.

\begin{table}[htbp]
	\centering
	\begin{scriptsize}
		\begin{tabular}{ c c c   }
			\hline\hline
			~~ &Exp&Fit  \\
			\hline
			
$\psi' \to \rho\pi$ & $0.032\pm0.012$& $0.037\pm0.010$ \\
$\psi' \to \rho^0\eta$ & $0.022\pm0.006$& $0.021\pm0.005$ \\
$\psi' \to \rho^0\eta'$ & $0.019\pm0.017$& $0.028\pm0.008$ \\

$\psi' \to \omega\pi^0$ & $0.021\pm0.006$& $0.021\pm0.004$ \\

$\psi' \to \omega\eta$ &---&  $0.005\pm0.003$\\
$\psi' \to \omega\eta'$ & $0.032\pm0.025$& $0.033\pm0.019$ \\

$\psi' \to \phi\eta$ & $0.031\pm0.0031$& $0.032\pm0.003$ \\
$\psi' \to \phi\eta'$ & $0.0154\pm0.0020$& $0.016\pm0.0019$ \\

$\psi' \to K^{*+}K^-+c.c.$ & $0.029\pm0.004$& $0.029\pm0.004$ \\
$\psi' \to K^{*0}\bar{K}^0+c.c.$ & $0.109\pm0.020$& $0.080\pm0.011$ \\		
			\hline\hline
		\end{tabular}
	\end{scriptsize}
	\caption{ Branching fractions$(\times 10^{-3})$ of the decay processes for $\psi'$. The $\psi'\to\omega\eta$ channel is not included in the fit, instead the result corresponds to our prediction, which is around two times smaller than the upper limit $1.1\times 10^{-5}$ reported in PDG~\cite{ParticleDataGroup:2022pth}.  \label{tab.Branching fractions for psip decays}  } 
\end{table}
 
With the fitted parameters in Table~\ref{tab.fitting}, it is then interesting to decipher the roles of different mechanisms and resonances played in a given process. 

The $\jpsi\to P l^+l^-$ processes can provide an environment to study the intermediate hadron resonances~\cite{Chen:2014yta,Kubis:2014gka}. Recently, the $\jpsi\to \eta \gamma^{*} (\to e^+e^-)$ form factors are reported by the BESIII collaboration in Ref.~\cite{BESIII:2018qzg}, in which the experimental analysis includes only the $\rho$ resonance in the $e^+e^-$ spectra, apart from the QED contributions. However, it is pointed out that the $\rho$ contribution should come from an isospin violated intermediate process $\jpsi\to \eta \rho \to \eta e^+e^-$. In contrast, the contributions from the $\omega$ and $\phi$ are expected to be more important, since they enter via the isospin conserved intermediate processes $\jpsi\to \eta \omega$ and $\jpsi\to \eta \phi$, whose branching ratios are around eight and four times larger than that of the $\jpsi\to\rho\eta$ in order. As a result, we expect that the effect of the $\rho$ resonance is much suppressed, compared to the contributions from $\omega$ and $\phi$. Due to the narrow widths of the latter two resonances, they manifest themselves as prominent peaks in the $e^+e^-$ spectra, as shown in Fig.~\ref{fig.dBdqFFpsietaee}. However, these narrow peaks can be easily washed out when the energy resolution is low. E.g., we also explicitly give the energy  distributions of the $e^+e^-$ in Fig.~\ref{fig.dBdqFFpsietaee} when taking the energy bin width at 50~MeV and 100~MeV. In the latter case the signals of the narrow $\omega$ and $\phi$ become faintly visible. As pointed out in Refs.~\cite{Kuang-TaChao:NPB1990,Q. Zhao:PLB2011,Chen:2014yta}, we also confirm the importance of the $\eta^{(')}-\eta_c$ mixing mechanism in the $\jpsi\to\eta^{(')}\gamma^{(*)}$ decay processes. A future experimental measurement with higher energy resolution will be definitely helpful to discriminate the roles of different hadrons in the $\jpsi\to \eta e^+e^-$ process. In Table~\ref{tab.Branching fractions for jpsi Pll}, we give our predictions to the branching ratios of various $\jpsi\to P l^+l^-$ processes and also make comparisons with the results in Refs.~\cite{Chen:2014yta,J. Fu:MPLA2012,He:2020jvj}.

\begin{table}[htbp]
	\centering
	\begin{scriptsize}
		\begin{tabular}{ c c c c c c}
			\hline\hline
		
		~~~~&Exp& This Work& Ref.\cite{Chen:2014yta} & Ref.\cite{J. Fu:MPLA2012} & Ref.\cite{He:2020jvj}\\
            
            \hline		
				
		$\psi \to \pi^0e^+e^-$ & $0.076\pm0.014$&$0.1294\pm0.0044$ &$0.1191\pm0.0138$&$0.0389^{+0.0037}_{-0.0033}$&—— \\
		
		$\psi \to \eta e^+e^-$ & $1.42\pm0.08$&$1.35\pm0.02$ &$1.16\pm0.08$&$1.21\pm0.04$&1.38 \\
		
		$\psi \to \eta' e^+e^-$ & $6.59\pm0.18$&$6.08\pm0.05$ &$5.76\pm0.16$&$5.66\pm0.16$&6.06 \\
		
		$\psi \to \pi^0\mu^+\mu^-$ & —— &$0.0304\pm0.0010$&$0.0280\pm0.0032$&$0.0101^{+0.0010}_{-0.0009}$&—— \\
		
		$\psi \to \eta \mu^+\mu^-$ & ——&$0.40\pm0.01$&$0.32\pm0.02$&$0.30\pm0.01$&0.46 \\
		
		$\psi \to \eta' \mu^+\mu^-$ &—— &$1.64\pm0.02$&$1.46\pm0.04$&$1.31\pm0.04$&1.72 \\
	\hline\hline
\end{tabular}
\end{scriptsize}
\caption{Branching ratios $(\times 10^{-5})$ for $J/\psi \to Pl^+l^-$ . \label{tab.Branching fractions for jpsi Pll}  }
\end{table}

\begin{table}[htbp]
	\centering
	\begin{scriptsize}
		\begin{tabular}{ c c c c }
			\hline\hline

			Isospin conserved cases &Exp & Strong interaction & EM interaction \\
			\hline		
			
		$\mid G_{J/\psi \to \rho^0\pi^0}\mid $&$2.537\pm0.154$& $2.899\pm0.075$&$0.385 \pm 0.006$\\
		$\mid G_{J/\psi \to \rho\pi}\mid $&$4.408\pm0.191$& $5.022\pm0.129$&$0.709 \pm 0.009$\\
		$\mid G_{J/\psi \to \omega\eta} \mid$&$1.497\pm0.084$& $1.586\pm0.037$&$0.132\pm 0.009$\\
		$\mid G_{J/\psi \to \omega\eta'} \mid$&$0.562\pm0.026$& $0.647\pm0.028$&$0.119\pm 0.008$\\
		$\mid G_{J/\psi \to \phi\eta} \mid$&$1.060\pm0.056$& $1.270\pm0.045$&$0.198\pm 0.031$\\
		$\mid G_{J/\psi \to \phi\eta'} \mid$&$0.974\pm0.052$& $1.074\pm0.049$&$2.031\pm 0.044$\\
		$\mid G_{J/\psi \to K^{*+}K^-} \mid$&$2.011\pm0.161$& $2.313\pm0.048$&$0.216 \pm 0.036$\\
		$\mid G_{J/\psi \to K^{*0}\bar{K}^0} \mid$&$1.686\pm0.078$& $2.308\pm0.048$&$0.715 \pm 0.009$\\
			
			\hline\hline
			
			Isospin violated cases&Exp & EM interaction &  Strong interaction  \\ \hline
			
			$\mid G_{J/\psi \to \rho^0\eta}\mid $&$0.498\pm0.029$&$0.487\pm0.028$& $--$\\
			$\mid G_{J/\psi \to \rho^0\eta'} \mid$&$0.367\pm0.018$&$0.365\pm0.017$& $--$\\
			$\mid G_{J/\psi \to \omega\pi^0} \mid$&$0.721\pm0.039$&$0.720\pm0.035$& $--$\\
			
			\hline\hline
			
		\end{tabular}
	\end{scriptsize}
	\caption{ The effective couplings of $G_{\psi \to VP}$ in units of $10^{-6}{\rm MeV}^{-1}$. \label{tab.The modulus of from factor for psiVP}  }
\end{table}

\begin{table}[htbp]
	\centering
	\begin{scriptsize}
		\begin{tabular}{ c c c c }
			\hline\hline			
			
			Isospin conserved cases&Exp & Strong interaction & EM interaction  \\
			\hline		
			
			$\mid G_{\psi'} \to \rho\pi\mid $&$0.255\pm0.044$&$0.029\pm0.036$&$0.255 \pm 0.016$\\
			$\mid G_{\psi'} \to \omega\eta \mid$&---&$0.103\pm0.038$&$0.036\pm 0.003$\\
			$\mid G_{\psi'} \to \omega\eta' \mid$&$0.288\pm0.096$&$0.212\pm0.079$&$0.079 \pm 0.013$\\
			$\mid G_{\psi'} \to \phi\eta \mid$&$0.275\pm0.013$&$0.285\pm0.030$&$0.565\pm 0.029$\\
			$\mid G_{\psi'} \to \phi\eta' \mid$&$0.213\pm0.013$&$0.197\pm0.068$&$0.412 \pm 0.067$\\
			$\mid G_{\psi'} \to K^{*+}K^-\mid $&$0.181\pm0.012$&$0.263\pm0.021$&$0.093 \pm 0.021$\\
			$\mid G_{\psi'} \to K^{*0}\bar{K}^0 \mid$&$0.352\pm0.031$&$0.267\pm0.021$&$0.568\pm 0.007$\\
			
			\hline\hline
			
			Isospin violated cases&Exp & EM interaction & Strong interaction  \\ \hline
			
			$\mid G_{\psi'} \to \rho^0\eta \mid$&$0.219\pm0.028$&$0.216\pm0.026$& $--$\\
			$\mid G_{\psi'} \to \rho^0\eta' \mid$&$0.222\pm0.083$&$0.271\pm0.040$& $--$\\
			$\mid G_{\psi'} \to \omega\pi^0 \mid$&$0.207\pm0.028$&$0.208\pm0.019$& $--$\\
			\hline\hline
			
		\end{tabular}
	\end{scriptsize}
	\caption{ The effective couplings of $G_{\psi' \to VP}$ in units of $10^{-6}{\rm MeV}^{-1}$. \label{tab.The modulus of from factor for psipVP}  }
\end{table}

Our study reveals an interesting feature that can shed light on the $\rho\pi$ puzzle in the $\jpsi(\psi')\to VP$ decays. For this purpose, let's focus on the interplay between the electromagnetic and strong interactions in the $\jpsi(\psi') \to V P$ processes. We separately show the contributions from the strong and electromagnetic interactions to the isospin conserved and violated decays for $\jpsi$ and $\psi'$ in Tables~\ref{tab.The modulus of from factor for psiVP} and ~\ref{tab.The modulus of from factor for psipVP}, respectively. The contributions from the strong interactions are given by the $h_{i=1,2,3}$ terms in Eq.~\eqref{eq.lagpsivp}, while the electromagnetic contributions are obtained by taking $h_{i=1,2,3}=0$. 
For the $\jpsi\to VP$ decays, the contributions from strong interactions turn out to play major roles in most of the isospin conserved channels, with the exception of the $\jpsi\to\phi\eta'$ process, where the strengths of the two types of interactions are comparable. While the isospin violated channels can only receive contributions from the electromagnetic interactions, since the isospin breaking effects from the strong interaction parts are not included in this work. According to the results shown in Table~\ref{tab.The modulus of from factor for psipVP}, for the $\psi' \to VP$ processes, the strong interactions are found to play comparable roles in many of the isospin conserved channels as those from the electromagnetic parts. Especially, the electromagnetic interaction turns out to play the dominant role in the $\psi'\to \rho \pi$ process and the effects from the strong interactions are found to be very small. In contrast, the strong interactions dominate the decay of $\jpsi\to \rho \pi$ process and the electromagnetic effects appear to be small. This provides a sensible explanation to the $\rho\pi$ puzzle.

For the charged $K^{*+}K^- + c.c. $ and neutral $K^{*0}\bar{K}^0 + c.c.$ decay processes of $\jpsi$ or $\psi'$, the $SU(3)$ breaking effects can originate from the strong interactions via the $h_2$ term in Eq.~\eqref{eq.lagpsivp}, which turns out to be the same for both charged and neutral processes, and the electromagnetic interactions via the $c_j$ terms in Eq.~\eqref{eq.lagvjp}, where the $c_4$ operator is found to solely contribute to the charged process~\cite{Chen:2013nna}. 
The contributions from the electromagnetic parts to the $\jpsi\to K^*\bar{K}+ c.c.$ processes are obviously smaller than those from the strong interactions, which explains the similar branching ratios between $\jpsi\to K^{*+}K^- + c.c. $ and $\jpsi\to K^{*0}\bar{K}^0 + c.c.$. In contrast, our study reveals that the magnitudes of the strong interactions in the $\psi' \to K^*\bar{K}+ c.c.$ can be comparable with those of the electromagnetic parts. While, the $SU(3)$ breaking effects in the electromagnetic parts are quite different for the charged and neutral decay processes due to the $c_4$ operator~\cite{Chen:2013nna}. This gives a new insight and also a reasonable explanation to the very different branching ratios of the $\psi'\to K^{*+}K^- + c.c. $ and $\psi'\to K^{*0}\bar{K}^0 + c.c.$.

\section{Summary and conclusions}\label{sec.summary}

We use the effective Lagrangian approach to simultaneously investigate the  processes of $\jpsi(\psi’) \to VP$, $\jpsi \to P\gamma$,$\jpsi \to Pl^+l^-$, the radiative decays of light-flavor hadrons and their relevant form factors. High energy constraints on the resonance couplings are used to reduce the number of free parameters. The remaining resonance couplings are then determined through the joint fit to a large amount of experimental data, including the updated PDG averages of the various partial decay widths and the most recent $J/\psi \to \eta \gamma^*$ form factors from BESIII.  

Thanks to the use of effective Lagrangian, the different types of contributions from the OZI allowed/suppressed strong interactions, $SU(3)$ breaking terms and  electromagnetic effects can be easily identified in our study. 
We pay special attention to the relative magnitudes from the strong and  electromagnetic interactions in the $\jpsi \to VP$ and $\psi’\to VP$ processes, so as to provide an insight into the $\rho\pi$ puzzle. An anatomy of the $\jpsi\to\rho \pi$ and $\psi'\to \rho \pi$ amplitudes reveals that the strong interaction dominates the former process and the electromagnetic interaction prevails the latter one. For the obviously distinct ratios between the charged $B(\psi'\to K^{*+}K^{-}+c.c.)/B(\jpsi\to K^{*+}K^{-}+c.c.)$ and the neutral $B(\psi'\to K^{*0}\bar{K}^{0}+c.c.)/B(\jpsi\to K^{*0}\bar{K}^{0}+c.c.)$ processes, our study uncovers that the $\jpsi\to K^{*}\bar{K}+c.c.$ processes are mainly ruled by the strong interactions, where the $SU(3)$ breaking effects enter similarly in both the charged and neutral amplitudes, while the $\psi'\to K^{*}\bar{K}+c.c.$ decays are found to be importantly affected by the electromagnetic interactions, where the $SU(3)$ symmetry breaking terms appear differently in the charged and neutral processes.

\section*{Acknowledgements}
We thank Lu Niu for an early-stage contribution to this work. 
This work is partially funded by the Natural Science Foundation of China under Grant Nos.~11975090, 12150013, 11975028 and 11974043.

\appendix 

\section{ The expressions of the effective couplings for $\psi' \to VP$}
\setcounter{equation}{0}
\def\theequation{\Alph{section}.\arabic{equation}}\label{appendix.FFpsip}

The expressions of the effective couplings in the $\psi' \to VP$ processes defined in Eq.~\eqref{widthpsipVP} take the form: 

\begin{align}
	G_{\psi^\prime \to \rho^0\pi^0}=&\frac{2\sqrt{2}}{F_\pi M_\rho} h^\prime _1 M_{\psi^\prime}+ \frac{8\sqrt{2}}{F_\pi M_\rho}h^\prime _2 m^2_\pi \frac{1}{M_{\psi^\prime}}+ \frac{32\pi\alpha}{F_\pi M_\rho}F_V g^\prime _1+ \frac{128\pi\alpha}{F_\pi M_\rho}F_V g^\prime _2 \frac{m^2_\pi}{M^2_{\psi^\prime}} \nonumber\\ &   +\frac{8\sqrt{2}\pi\alpha}{3}\frac{f_{\psi^\prime}}{M_{\psi^\prime}}F_{\rho\pi\gamma^*}(M^2_{\psi^\prime})\,,
\end{align}

\begin{align}
	G_{\psi^\prime \to \rho^+\pi^-}=&\frac{2\sqrt{2}}{F_\pi M_\rho} h^\prime _1 M_{\psi^\prime}+ \frac{8\sqrt{2}}{F_\pi M_\rho}h^\prime _2 m^2_\pi \frac{1}{M_{\psi^\prime}}+ \frac{8\sqrt{2}\pi\alpha}{3}\frac{f_{\psi^\prime}}{M_{\psi^\prime}}F_{\rho\pi\gamma^*}(M^2_{\psi^\prime})\,,
\end{align}

\begin{align}
	G_{\psi^\prime \to \rho^0\eta}=&\frac{32\sqrt{2}\pi\alpha}{3FM_\rho}F_V g^\prime _1(a_1-a_3)+\frac{128\sqrt{2}\pi\alpha}{3FM_\rho M^2_{\psi^\prime}}F_V g^\prime _2 [a_1m^2_\pi-a_3(2m^2_K-m^2_\pi)]\notag\\
	&-8\pi\alpha\frac{F_V}{M_\rho} \lambda_{\eta\eta_c} g_{\psi^\prime\eta_c\gamma^{*}}(M^2_\rho)e^{i\delta^\prime _\eta}+\frac{8\sqrt{2}\pi\alpha}{3} \frac{f_{\psi^\prime}}{M_{\psi^\prime}} F_{\rho\eta\gamma^*}(M^2_{\psi^\prime})\,,
\end{align}

\begin{align}
	G_{\psi^\prime \to \rho^0\eta^\prime}=&\frac{32\sqrt{2}\pi\alpha}{3FM_\rho}F_V g^\prime _1(a_2-a_4)+\frac{128\sqrt{2}\pi\alpha}{3FM_\rho M^2_{\psi^\prime}}F_V g^\prime _2 [a_2m^2_\pi-a_4(2m^2_K-m^2_\pi)]\notag\\
	&-8\pi\alpha\frac{F_V}{M_\rho} \lambda_{\eta^\prime\eta_c} g_{\psi^\prime\eta_c\gamma^{*}}(M^2_\rho)e^{i\delta^\prime _{\eta^ \prime}}+\frac{8\sqrt{2}\pi\alpha}{3} \frac{f_{\psi^\prime}}{M_{\psi^\prime}} F_{\rho\eta^\prime\gamma^*}(M^2_{\psi^\prime})\,,
\end{align}

\begin{align}
	G_{\psi^\prime \to \omega\pi^0}=&\frac{32\pi\alpha}{3F_\pi M_\omega} F_V g^\prime _1+\frac{128\pi\alpha}{3F_\pi M_\omega} F_V g^\prime _2
	\frac{m^2_\pi}{M^2_{\psi^\prime}}+\frac{8\sqrt{2}\pi\alpha}{3}\frac{f_{\psi^\prime}}{M_{\psi^\prime}}F_{\omega\pi\gamma^*}(M^2_{\psi^\prime})\,,
\end{align}

\begin{align}
	G_{\psi^\prime \to \omega\eta}=&\frac{4}{F M_\omega}a_1 h^\prime_1M_{\psi^\prime}+\frac{16}{F M_\omega} a_1 h^\prime_2 m^2_{\pi} \frac{1}{M_{\psi^\prime}} +\frac{4}{F M_\omega}(2a_1+a_3)h^\prime_3 M_{\psi^\prime}+\frac{32\sqrt{2}\pi\alpha}{9FM_\omega}F_V g^\prime_1(a_1-a_3)\notag\\
	&+\frac{128\sqrt{2}\pi\alpha}{9FM_\omega M^2_{\psi^\prime}}F_V g^\prime _2 [a_1m^2_\pi-a_3(2m^2_K-m^2_\pi)]-\frac{8}{3}\pi\alpha\frac{F_V}{M_\omega} \lambda_{\eta\eta_c} g_{\psi^\prime\eta_c\gamma^{*}}(M^2_\omega)e^{i\delta^\prime _{\eta}}\notag\\
	&+\frac{8\sqrt{2}\pi\alpha}{3} \frac{f_{\psi^\prime}}{M_{\psi^\prime}} F_{\omega\eta\gamma^*}
	(M^2_{\psi^\prime})\,,
\end{align}

\begin{align}
	G_{\psi^\prime \to \omega\eta^\prime}=&\frac{4}{F M_\omega}a_2 h^\prime_1M_{\psi^\prime}+\frac{16}{F M_\omega} a_2 h^\prime_2 m^2_{\pi} \frac{1}{M_{\psi^\prime}} +\frac{4}{F M_\omega}(2a_2+a_4)h^\prime_3 M_{\psi^\prime}+\frac{32\sqrt{2}\pi\alpha}{9FM_\omega}F_V g^\prime_1(a_2-a_4)\notag\\
	&+\frac{128\sqrt{2}\pi\alpha}{9FM_\omega M^2_{\psi^\prime}}F_V g^\prime _2 [a_2m^2_\pi-a_4(2m^2_K-m^2_\pi)]-\frac{8}{3}\pi\alpha\frac{F_V}{M_\omega} \lambda_{\eta^\prime\eta_c} g_{\psi^\prime\eta_c\gamma^{*}}(M^2_\omega)e^{i\delta^\prime _{\eta^ \prime}}\notag\\
	&+\frac{8\sqrt{2}\pi\alpha}{3} \frac{f_{\psi^\prime}}{M_{\psi^\prime}} F_{\omega\eta^\prime\gamma^*}
	(M^2_{\psi^\prime})\,,
\end{align}

\begin{align}
	G_{\psi^\prime \to \phi\eta}=&-\frac{2\sqrt{2}}{F M_\phi}a_3 h^\prime_1 M_{\psi^\prime}-\frac{8\sqrt{2}}{F M_\phi} a_3 h^\prime_2
	(2m^2_K-m^2_\pi)\frac{1}{M_{\psi^\prime}}-\frac{2\sqrt{2}}{F M_\phi}(2a_1+a_3)h^\prime_3 M_{\psi^\prime} \notag\\
	&+\frac{64\pi\alpha}{9FM_\phi}F_V g^\prime _1(a_1-a_3)+\frac{256\pi\alpha}{9FM_\phi M^2_{\psi^\prime}}F_V g^\prime _2 [a_1m^2_\pi-a_3(2m^2_K-m^2_\pi)]\notag\\
	&-\frac{8\sqrt{2}}{3M_\phi}\pi\alpha F_V \lambda_{\eta\eta_c} g_{\psi^\prime\eta_c\gamma^{*}}(M^2_\phi)e^{i\delta^\prime _\eta}+\frac{8\sqrt{2}\pi\alpha}{3}\frac{f_{\psi^\prime}}{M_{\psi^\prime}} F_{\phi\eta\gamma^*}(M^2_{\psi^\prime})\,,
\end{align}

\begin{align}
	G_{\psi^\prime \to \phi\eta^\prime}=&-\frac{2\sqrt{2}}{F M_\phi}a_4 h^\prime_1 M_{\psi^\prime}-\frac{8\sqrt{2}}{F M_\phi} a_4 h^\prime_2
	(2m^2_K-m^2_\pi)\frac{1}{M_{\psi^\prime}}-\frac{2\sqrt{2}}{F M_\phi}(2a_2+a_4)h^\prime_3 M_{\psi^\prime}\notag \\
	&+\frac{64\pi\alpha}{9FM_\phi}F_V g^\prime _1(a_2-a_4)+\frac{256\pi\alpha}{9FM_\phi M^2_{\psi^\prime}}F_V g^\prime _2 [a_2m^2_\pi-a_4(2m^2_K-m^2_\pi)]\notag\\
	&-\frac{8\sqrt{2}}{3M_\phi}\pi\alpha F_V \lambda_{\eta^\prime\eta_c} g_{\psi^\prime\eta_c\gamma^{*}}(M^2_\phi)e^{i\delta^\prime _{\eta^ \prime}}+\frac{8\sqrt{2}\pi\alpha}{3}\frac{f_{\psi^\prime}}{M_{\psi^\prime}} F_{\phi\eta^\prime\gamma^*}(M^2_{\psi^\prime})\,,
\end{align}

\begin{align}
	G_{\psi^\prime \to K^{*+}K^-}=\frac{2\sqrt{2}}{F_K M_{K^*}}h^\prime_1 M_{\psi^\prime}+\frac{8\sqrt{2}}{F_K M_{K^*}}h^\prime_2 m^2_K
	\frac{1}{M_{\psi^\prime}}+\frac{8\sqrt{2}\pi \alpha}{3}\frac{f_{\psi^\prime}}{M_{\psi^\prime}}F_{K^{*+}K^-\gamma^*}(M^2_{\psi^\prime})\,,
\end{align}

\begin{align}
	G_{\psi^\prime \to K^{*0}\bar{K}^0}=\frac{2\sqrt{2}}{F_K M_{K^*}}h^\prime_1 M_{\psi^\prime}+\frac{8\sqrt{2}}{F_K M_{K^*}}h^\prime_2 m^2_K
	\frac{1}{M_{\psi^\prime}}+\frac{8\sqrt{2}\pi \alpha}{3}\frac{f_{\psi^\prime}}{M_{\psi^\prime}}F_{K^{*0}\bar{K}^0\gamma^*}(M^2_{\psi^\prime})\,,
\end{align}
with 
\begin{align}
	a_1=&\frac{F}{\cos(\theta_0-\theta_8)}(\frac{\cos\theta_0}{\sqrt{6}F_8}-\frac{\sin\theta_8}{\sqrt{3}F_0})\,, \nonumber\\
	a_2=&\frac{F}{\cos(\theta_0-\theta_8)}(\frac{\sin\theta_0}{\sqrt{6}F_8}+\frac{\cos\theta_8}{\sqrt{3}F_0})\,, \nonumber\\
	a_3=&\frac{F}{\cos(\theta_0-\theta_8)}(\frac{-2\cos\theta_0}{\sqrt{6}F_8}-\frac{\sin\theta_8}{\sqrt{3}F_0})\,, \nonumber\\
	a_4=&\frac{F}{\cos(\theta_0-\theta_8)}(\frac{-2\sin\theta_0}{\sqrt{6}F_8}+\frac{\cos\theta_8}{\sqrt{3}F_0})\,. 
\end{align} 
The expressions $F_{K^{*+}K^-\gamma^*}$ and $F_{K^{*0}\bar{K}^0\gamma^*}$ that  correspond to the electromagnetic contributions to the effective couplings of $G_{\psi' \to K^{*+}K^-}$ and $G_{\psi' \to K^{*0}\bar{K}^0}$ are given by 
\begin{align}
	F_{K^{*+}K^-\gamma^*}(s)=&\frac{-2\sqrt{2}}{3F_KM_VM_{K^*}}[(c_1+c_2+8c_3-c_5)m^2_K+(c_2+c_5-c_1-2c_6)M^2_{K^*}+(c_1-c_2+c_5)s\notag\\
	&+24c_4(m^2_K-m^2_{\pi})]+\frac{2F_V}{3F_KM_{K^*}}[(d_1+8d_2-d_3)m^2_K+d_3(M^2_{K^*}+s)]\notag\\
	&\times [D_\omega(s)+3D_\rho(s)-2D_\phi(s)],
\end{align}
and 
\begin{align}
	F_{K^{*0}\bar{K}^0\gamma^*}(s)=&\frac{4\sqrt{2}}{3F_KM_VM_{K^*}}[(c_1+c_2+8c_3-c_5)m^2_K+(c_2+c_5-c_1-2c_6)M^2_{K^*}+(c_1-c_2+c_5)s]\notag\\
	&+\frac{2F_V}{3F_KM_{K^*}}[(d_1+8d_2-d_3)m^2_K+d_3(M^2_{K^*}+s)]  [D_\omega(s)-3D_\rho(s)-2D_\phi(s)]\,. 
\end{align}
It is pointed out that $SU(3)$ breaking effect caused by the $c_4$ term only enters in the charged $F_{K^{*+}K^-\gamma^*}$ amplitude and is absent in the neutral  $F_{K^{*0}\bar{K}^0\gamma^*}$ process. For the expressions of other amplitudes $F_{VP\gamma^{(*)}}$, they are given in Appendix of Ref.~\cite{Chen:2014yta} and we do not repeat them here.

\end{document}